\documentclass[doublecol]{epl2} 
\usepackage{amsmath,amsfonts} 
\usepackage{graphicx,color} 
\usepackage{psfrag}
\usepackage[normalem]{ulem}

\newcommand{\pc}{p_{\scriptstyle \rm c}}
\newcommand{\Bc}{B_{\scriptstyle \rm c}}

\newcommand{\erf}{{\rm erf}}
\newcommand{\erfc}{{\rm erfc}}

\date{\today}

\title{The large connectivity limit of bootstrap percolation}
\shorttitle{The large connectivity limit of bootstrap percolation}

\author{Giorgio Parisi\inst{1} \and Mauro Sellitto\inst{2}}

\institute{\inst{1}Dipartimento di Fisica, Universit\`a ``La
  Sapienza'' and INFN, Piazzale A. Moro 2, I-00185 Roma, Italy
  \\ \inst{2}Dipartimento di Ingegneria Industriale e
  dell'Informazione, Seconda Universit\`a di Napoli, 
Via Roma 29, I-81031 Aversa (CE), Italy}

\pacs{64.60.Kw}{Multicritical points}
\pacs{05.70.Fh}{Phase transitions: general studies}

\abstract{Bootstrap percolation provides an emblematic instance of
  phase behavior characterised by an abrupt transition with diverging
  critical fluctuations. This unusual hybrid situation generally
  occurs in particle systems in which the occupation probability of a
  site depends on the state of its neighbours through a certain
  threshold parameter. In this paper we investigate the phase behavior
  of the bootstrap percolation on the regular random graph in the
  limit in which the threshold parameter and lattice connectivity
  become both increasingly large while their ratio $\alpha$ is held
  constant. We find that the mixed phase behavior is preserved in this
  limit, and that multiple transitions and higher-order bifurcation
  singularities occur when $\alpha$ becomes a random variable.}

\begin{document}

\maketitle

\section{Introduction}

Bootstrap percolation (BP) is a generalization of the ordinary random
percolation problem to particle systems in which the occupation
probability of a site depends on the state of its
neighbours~\cite{Dawson}. It was introduced in the late 1970s in
connection with the behavior of some dilute magnets in which, under
certain circumstances, an atom displays a magnetic moment only if the
number of its magnetic neighbors is above a certain threshold.  Since
physical details of atoms and interactions are completely stripped
away, BP naturally lends itself to quite distinct possible
interpretations. Indeed, perhaps not surprisingly, threshold models of
cooperative behavior closely related to BP have been considered early
in the sociological literature~\cite{socio} and have been applied to a
wide range of problems (see, for a list of examples,
Ref.~\cite{Dawson}).

The solution of BP on the Bethe lattice~\cite{Chalupa} showed the
presence of a peculiar phase transition in which a sudden jump of the
order parameter is accompanied by diverging fluctuations as in
ordinary critical phenomena.  For this reason the BP transition is
said to have a mixed or hybrid nature.  Interest in mixed phase
behavior has grown in the last years while it has been recognized that
its occurrence in apparently unrelated problems, such as slow
relaxation in supercooled liquids~\cite{Goetze_book,WoLu_book},
jamming in granular materials~\cite{LiuNagel,PaZa}, and hardness in
combinatorial optimization problems~\cite{HaWe_book,MeMo_book} may
have a common microscopic origin~\cite{MoZe,SeBiTo05,ScLiCh06}, and be
generally relevant to the vulnerability of complex networks under
damage~\cite{Goltsev}.

In this paper, we consider the BP problem on the regular random graph
in the limit in which the threshold parameter and lattice connectivity
become both increasingly large while holding their ratio
constant. This type of mean-field limit is interesting because it
offers the possibility to get extra analytical insights into the
connectivity and threshold dependence of relevant quantities such as
the critical point and the jump of the order parameter at the
transition. We find, for example, that mixed phase behavior is
generally robust, and that the critical amplitude of the order
parameter decreases as a power-law of connectivity with an exponent of
$-1/4$. This latter feature implies that the infinite connectivity
limit is, somewhat surprisingly, singular. We also show that this
framework is general enough to deal with complex phase behaviors which
arise when the nodes of the random graph may take different values of
the ratio of threshold parameter to connectivity. Depending on the
relative fractions of the different type of nodes one can then observe
higher-order bifurcations and multiple phase
transitions~\cite{Branco,SeDeCaAr,Cellai,Porto,Se2012}, as we shall
see.

\section{The model}

Consider a lattice system in which every site is first randomly
occupied by a particle with probability $p$. Then, remove from the
lattice every particle which is surrounded by a number of neighbouring
particles less than $m$.  Iterating this procedure leads to two
possible asymptotic results~\cite{Chalupa}: either the lattice becomes
completely empty or there is a residual cluster of particles in which
every particle has at least $m$ neighboring particles. On tree-like
structure the residual cluster is infinite and the BP problem can be
solved exactly, in particular one can determine the initial critical
particle density $\pc$ above which a spanning cluster appears.  When
$m=1,\, 2$, random percolation (with and without dangling ends) is
recovered. In this case, the phase transition is continuous and the
structure of the incipient spanning cluster is fractal.  We focus here
on the case $m > 2$. We consider a Bethe lattice with fixed
connectivity, specifically, a regular random graph with coordination
number $z=k+1$.  Let us call $B$ the probability that a site is
connected to the infinite cluster through a nearest neighbour.  $B$
verifies the self-consistent equation~\cite{Chalupa}
\begin{eqnarray}
  B &=& p \ \sum_{i=m-1}^k {k \choose i} B^i (1-B)^{k-i} ,
\label{eq.B}
\end{eqnarray}
and in the following it will play the role of the order
parameter~\cite{NB}.  It is well known~\cite{Chalupa} that when $m >
2$ the trivial solution of Eq.~(\ref{eq.B}), $B=0$, becomes unstable
above a critical point $\pc$ that depends on both the connectivity and
threshold. The new solution emerges discontinuously at $\pc$ with a
jump that behaves near the transition point as $B-\Bc \sim
\sqrt{p-\pc}$, and therefore critical fluctuations of the order
parameter are expected from the divergence of $dB/dp$ when $p \to
\pc^{+}$. The geometric nature of critical fluctuations comes from the
divergence of the mean size of {\it corona} clusters, i.e., clusters
in which every particles is surrounded by exactly $m$ neighbouring
particles~\cite{Goltsev}. This marginality condition implies that the
corona structure is extremely fragile: when a corona particle is
removed the nearby particles become unstable under the BP rule. This
triggers a cascade process which, by a domino-like effect, leads to
the complete destruction of the corona~\cite{Goltsev}.

The BP can be generalised by attaching to every node of the random
graph its own threshold parameter. In this way one can explore a
larger variety of situations in which, e.g., particles may have
different sizes or shapes and, more generically, individuals are
characterised by distinct (threshold) behaviors.  Due to the
competition among clusters with more or less compact structures,
additional interesting features emerge in this heterogeneous case,
such as multicritical like point and multiple phase
transitions~\cite{Branco,SeDeCaAr,Cellai,Porto,Se2012}.  In the
following, we shall be essentially concerned with phase behavior of
mixed nature.

\begin{figure*}[htbp]
\begin{center}
\input{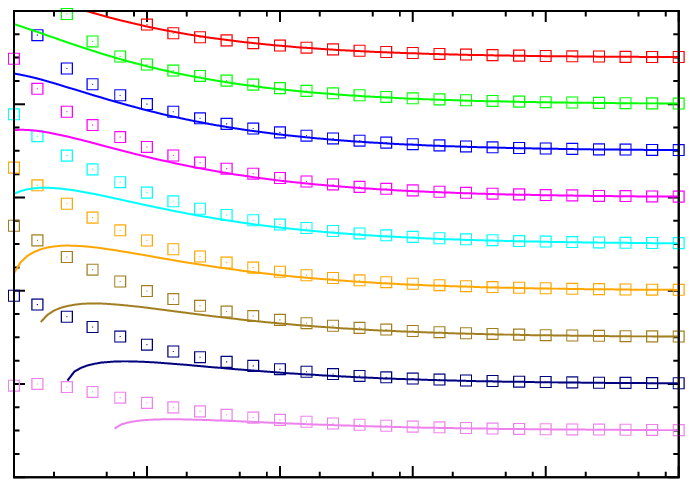}\input{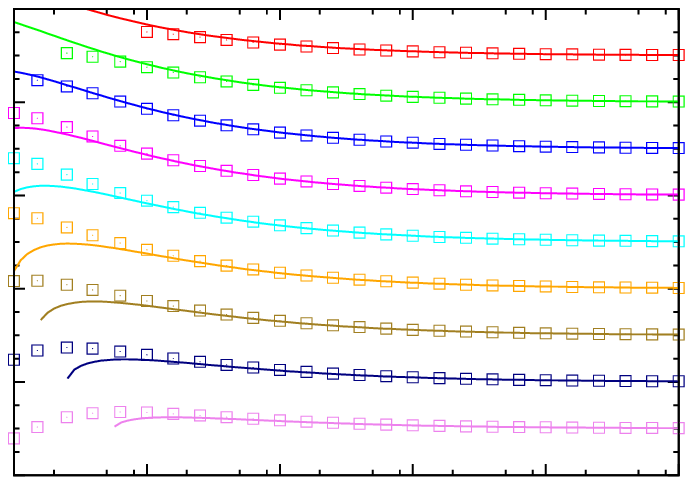}
\caption{Scaling of the critical point, $\pc$, and the order
  parameter, $\Bc$, with the random graph connectivity $k$ for
  $\alpha$ going from 0.1 to 0.9 with a step of 0.1 (from the bottom
  to the top). Square symbols denote the exact solution of
    Eq.~(\ref{eq_B}) while full lines represent the approximated
    analytical behavior given by Eqs.~(\ref{eq.Bc}) and
    (\ref{eq.pc}).}
\label{fig.pc_Bc}
\end{center}
\end{figure*}

\section{The large connectivity limit}

The binomial structure of Eq.~(\ref{eq.B}) leads naturally to the
possibility of considering the limit in which the threshold parameter
and lattice connectivity are both large, $m \gg1$ and $k \gg 1$, while
their ratio $\alpha=m/k$ is a constant, $0 < \alpha < 1$. In this
limit, according to the de Moivre-Laplace formula, each term
  of the sum on the right hand side of Eq.~(\ref{eq.B}) tends to a
  Gaussian probability density function:
\begin{eqnarray}
 G(i) = \frac{1}{\sqrt{2 \pi k B (1-B)}} \exp \left[ - \frac{(i-B
     k)^2}{ 2 k B (1-B)} \right],
\end{eqnarray}
with average $\mu=k \, B$ and variance $\sigma^2 = k \, B \, (1-B)$.
Passing to the continuous limit, the sum in Eq.~(\ref{eq.B}) is
converted to an integral and we can finally write down:
\begin{eqnarray}
  B &=& \frac{p}{2} \left( \erf \, Y_1 - \erf \, Y_{\alpha} \right) ,
\label{eq_B}
\end{eqnarray}
where $Y_1 = Y_{\alpha=1}$ and
\begin{eqnarray}
 Y_{\alpha}(B) = \frac{ \sqrt{k} (\alpha-B) }{ \sqrt{2B(1-B)} }.
\label{def_Y}
\end{eqnarray}
We first observe that Eq.~(\ref{eq_B}) has always the trivial solution
$B=0$ because the argument of the two error functions both diverge
(regardless the value of $k$). Additional solutions emerging
discontinuously at the critical point $\pc(k)$ are obtained as
follows.  For $k \to \infty$ we observe that the right hand side of
Eq.~(\ref{eq_B}) is either zero or $p$ depending on whether $B$ is
smaller or larger than $\Bc(\infty)=\alpha$. Therefore, we get for the
critical point $\pc(\infty) = \alpha$, and thus $B-\Bc(\infty) =
[p-\pc(\infty)]^{\beta}$, with $\beta=1$. As we shall see, this linear
behavior is however singular, i.e., it strictly holds only for
$k=\infty$.  For any large but finite $k$ the nature of the phase
transition has generally a mixed character, $B-\Bc(\alpha, k) \sim
[p-\pc(\alpha,k)]^{\beta}$ with $\beta=1/2$, over a sufficiently small
region near the critical point, provided that $\alpha$ is neither zero
nor one.

\subsection{The fold bifurcation}

To determine the location of the discontinuous critical point for
finite yet large $k$ we proceed as follows. We first rewrite the
self-consistent equation for $B$, eq.~(\ref{eq_B}) as
\begin{eqnarray}
\frac{1}{p} = {\mathcal F}(B)
\label{eq.p}
\end{eqnarray}
where
\begin{eqnarray}
  {\mathcal F}(B) = \frac{ \erf \, Y_1 - \erf \, Y_{\alpha} }{ 2 B }
  \simeq \frac{ \erfc \, Y_{\alpha} }{ 2 B } .
  \label{def_F}
\end{eqnarray}
Then, since we are seeking a discontinuous solution with critical
fluctuations (i.e., with $\beta<1$), we set ${\mathcal F}'(B) \sim
dp/dB$ to zero to get:
\begin{eqnarray}
  \erfc \, Y_{\alpha} & \simeq & - \frac{ 2 B }{ \sqrt{\pi} }
  \frac{dY_{\alpha}}{dB} \exp(-Y_{\alpha}^2) .
  \label{eq_F1}
\end{eqnarray}
To find a closed expression for $\Bc(\alpha,k)$ we 
approximate
\begin{eqnarray}
 Y_{\alpha} \simeq \frac{ \sqrt{k} (\alpha - B)}{
   \sqrt{2\alpha(1-\alpha)} }, \qquad \frac{dY_{\alpha}}{dB} \simeq
 \frac{- \sqrt{k}}{ \sqrt{2\alpha(1-\alpha)} },
\end{eqnarray}
and 
use the asymptotic expansion of the complementary error
function for large negative values of its argument:
\begin{eqnarray}
  \erfc(x) & \simeq & 2 + \frac{ {\rm e}^{-x^2} }{ x \sqrt{\pi} } .
\label{eq.erfc}
\end{eqnarray}
Plugging this expression in Eq.~(\ref{eq_B}) we obtain to the leading
order in $k$:
\begin{eqnarray}
  \Bc(\alpha, k) & \simeq & \alpha - \sqrt{ \frac{\alpha(1-\alpha)
    }{k} \ln{ \frac{k \, \alpha}{2 \pi (1-\alpha)} } }.
\label{eq.Bc}
\end{eqnarray}
Correspondingly, in this approximation the phase transition point is
located at
\begin{eqnarray}
\pc(\alpha,k) & \simeq & \Bc(\alpha,k).
\label{eq.pc}
\end{eqnarray}
In the Fig.~\ref{fig.pc_Bc} Eqs.~(\ref{eq.Bc}) and (\ref{eq.pc}), are
compared with the exact solution as a function of $k$ and for several
values of $\alpha$. The asymptotic approximation appears to work
pretty well when $k$ is generally larger than a few hundreds. In this
region $\Bc(\alpha,k)/\alpha$ depends solely on the scaled variable $k
\, \alpha/(1-\alpha) \simeq \mu^2/\sigma^2$, as can be easily argued
from Eq.~(\ref{eq.Bc}).

\begin{figure}[htbp]
\input{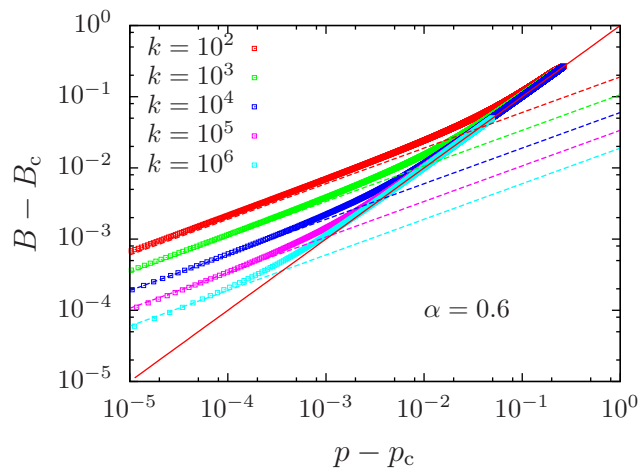}
\caption{Order parameter jump, $B-\Bc$, vs distance to the critical
  point, $p-\pc$, for several values of $k$, with $\alpha=0.6$.  The
  dashed straight lines are proportional to $k^{-1/4}
  \sqrt{p-\pc}$. The full line with slope one represents the linear
  behavior regime.}
\label{fig.B_vs_alpha04}
\end{figure}

To find the behavior of the $B(p)$ near the critical point we take the
derivative of the Eq.~(\ref{eq.p}):
\begin{eqnarray}
  \frac{dp}{dB}
  & \sim & \sqrt{k} (B-\Bc).
\end{eqnarray}
Close enough to the critical point,
\begin{eqnarray}
  p-\pc & \sim & \sqrt{k} (B-\Bc)^2,
\end{eqnarray}
which finally gives, to the leading order in $k$:
\begin{eqnarray}
  B-\Bc & \sim & k^{-1/4} \sqrt{p-\pc}.
\end{eqnarray}
We obtain therefore the well known square-root dependence of the order
parameter jump from the distance to the transition point which
characterises the singular behavior near a fold bifurcation. Notice
that in our mean-field limit the critical amplitude of the jump
decrease as a power-law of the connectivity with an exponent 1/4.
These behaviors perfectly agree with the exact solution of
Eq.~(\ref{eq_B}) as shown in the Fig.~\ref{fig.B_vs_alpha04}. When we
are sufficiently deep in the critical region the order parameter
depends linearly on the distance from the critical point, see
Fig.~\ref{fig.B_vs_alpha04}.  Comparing the linear and the square-root
dependence one can thus find the crossover point between these two
behaviors and get that the range of the square-root dependence
decreases with the connectivity as $k^{-1/2}$.

\subsection{Multiple transitions and higher bifurcations}

The fold bifurcation discussed above, which is denoted with ${\mathsf
  A}_{2}$ in the Arnold's terminology, represents only the lowest
level of the hierarchy of bifurcation singularities~\cite{Arnold}.  It
has been recently observed that the formal structure of the
self-consistent equation obeyed by $B$ in BP is essentially similar to
that satisfied by the nonergodicity parameter in the mode-coupling
theory of the liquid-glass transition~\cite{Se2012}. Consequently, one
can devise suitable generalizations of the BP problem which provide a
microscopic realization of the hierarchy of bifurcations in close
analogy with that appearing in some glassy systems~\cite{Se2012}.

We remind that an ${\mathsf A}_{\ell}$ bifurcation of order $\ell$
corresponds to the maximum root of Eq.~(\ref{eq.p}) which
simultaneously satisfies the equations:
\begin{equation}
\frac{d^n {\mathcal F}}{dB^n} = 0 \,,\,\,\, n=1, \dots, \ell-1; \qquad
\frac{d^{\ell} {\mathcal F}}{dB^{\ell}} \neq 0.
\label{eq.dF}
\end{equation}
Bifurcation singularities are interesting because they have distinct
critical properties. For example, the Taylor expansion of ${\mathcal
  F}$ near the bifurcation surface and Eqs.~(\ref{eq.dF}), imply that
the order parameter scales like $(p-\pc)^{1/\ell}$ near an ${\mathsf
  A}_{\ell}$ singularity. This means that order parameter fluctuations
are stronger when the order of the bifurcation singularity increases.

A generalization of BP in which higher-order bifurcations and multiple
transitions are possible can be obtained by letting the $\alpha$
parameter depend on the lattice sites. Notice that in the presence of
multiple roots of Eq.~(\ref{eq.p}) one has always to select the
maximum one because $B$ is a monotonically increasing function of
$p$. Here we consider the case in which $\alpha$ is randomly
distributed as
\begin{eqnarray}
{\mathcal P}(\alpha) = \sum_{i=1}^n q_i \delta(\alpha-\alpha_i) ,
\qquad \sum_{i=1}^n q_i =1,
\end{eqnarray}
with $q_i$ being the fraction of sites having threshold $\alpha_i$.
In this situation the equation for $B$ becomes
\begin{eqnarray}
  B &=& \frac{p}{2} \left( \erf \, Y_1 - \sum_{i=1}^n q_i \, \erf \,
  Y_{\alpha_i} \right) .
\label{eq_B_12}
\end{eqnarray}
Hereafter, we focus exclusively on binary mixtures of threshold
parameters, $n=2$, with $0<\alpha_i<1$. We note, however, that it is a
simple matter to extend our calculations to $n$-ary mixtures and to
situations in which one of the threshold parameters approaches one.
In the latter limit one recovers the ordinary random percolation which
may be interesting for those systems in which the interplay of
continuous and discontinuous percolation transitions is relevant.

\subsection{Infinite connectivity}

\begin{figure*}[htbp]
\begin{center}
{\bf a)}\input{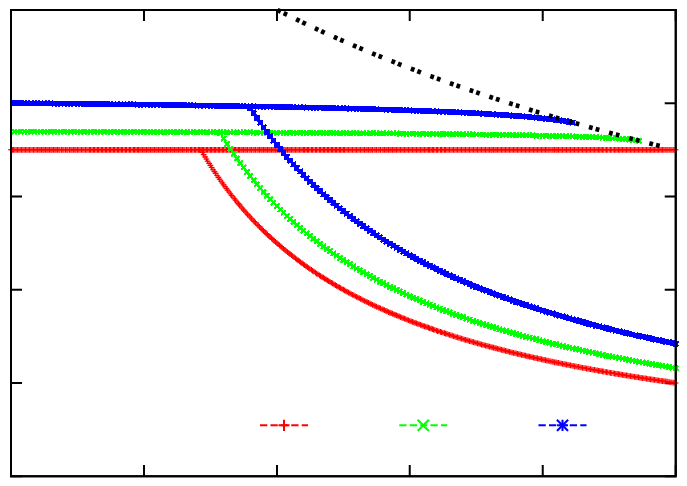}{\bf b)}\input{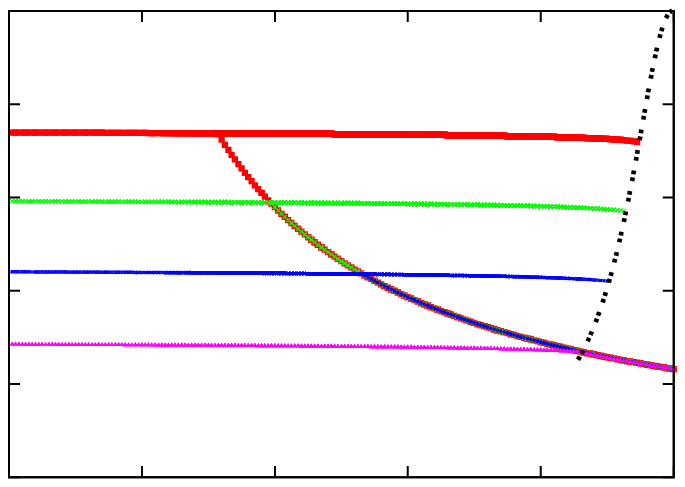}
\caption{Two sections of the phase diagram of generalised BP in the
  control variables $p$ and $q_2$.  The double-dotted lines in either
  figures represent cuts in the surface of ${\mathsf A}_3$
  singularities.  {\bf a)} $\alpha_1=0.7$ and $\alpha_2=0.2$ for three
  values of $k$. {\bf b)} $k=10^3$, $\alpha_2=0.2$ and
  $\alpha_1=0.7,\, 0.55,\, 0.4,\, 0.25$ from top to bottom; the
  ${\mathsf A}_4$ bifurcation corresponds to the terminal point of the
  double-dotted line.}
\label{fig.phase_gg}
\end{center}
\end{figure*}

We now first look at the phase structure of a binary mixture in the
limit $k \to \infty$ which will provide a guideline for the case of
large $k$. We use $p$ and $q_2$ as control variables.  Without loss of
generality we assume for sake of simplicity that $\alpha_1 \ge
\alpha_2$, and distinguish three regions for the order parameter.
\begin{itemize}
\item[$i)$] When $\alpha_1 \ge \alpha_2 \ge B$ one has $\erf \,
  Y_{\alpha_1} = \erf \, Y_{\alpha_2} = 1$ and thus $B=0$, which
  corresponds to the empty phase where every particle of the initial
  configuration can be removed by using the BP rule.

\item[$ii)$] For $B \ge \alpha_1 \ge \alpha_2$ we have $\erf \,
  Y_{\alpha_1} = \erf \, Y_{\alpha_2} = -1 $ which implies $B=p>0$
  meaning that there is a finite fraction of particles which cannot be
  removed from the system.  The phase boundary is obtained when
  $B=\alpha_1$, and is therefore a horizontal line in the plane
  $(p,q_2)$ located at
  \begin{eqnarray}
    p_{\scriptstyle \rm c_1} & =& \alpha_1.
    \label{eq.pc1_oo}
  \end{eqnarray} 
  This transition line is properly defined in the range $q_2 \in
  [0,1)$, because in the limit $q_2 \to 1$ we have to recover
    correctly the pure BP problem which has a transition located at
    $p_{\scriptstyle \rm c_2}=\alpha_2$ (as the fraction $q_1$ of
    type-1 sites is zero). The region $p \ge p_{\scriptstyle \rm c_1}$
    will be denoted as BP$_1$ phase and is depicted in the
    Fig.~\ref{fig.phase_gg}a.

\item[$iii)$] Finally, when $\alpha_1 \ge B \ge \alpha_2$ we get $\erf
  \, Y_{\alpha_1} = 1$ and $\erf \, Y_{\alpha_2} = -1 $ and therefore
  $B = p q_2$. The phase boundary with the empty phase corresponds to
  $B= \alpha_2$, and is located at:
  \begin{eqnarray}
    p_{\scriptstyle \rm c_2} & = & \frac{\alpha_2}{q_2} . 
   \label{eq.pc2_oo}
  \end{eqnarray}
  The region immediately above the hyperbolic branch,
  Eq.~(\ref{eq.pc2_oo}), will be denoted as BP$_2$ phase, see
  Fig.~\ref{fig.phase_gg}a. This loose phase is intermediate between
  the empty phase and the more compact BP$_1$ one: it appears as soon
  as the fraction $q_2$ of type-2 sites, becomes sufficiently large to
  support a percolating structure.  Notice that the stability limit of
  Eq.~(\ref{eq.pc2_oo}) lies above the crossover point with the first
  transition line, Eq~(\ref{eq.pc1_oo}), which is given by $
  p_{\scriptstyle \rm c_1} = p_{\scriptstyle \rm c_2}$, that is:
 \begin{eqnarray}
    q_{cr} &=& \frac{\alpha_2}{\alpha_1} ,
  \end{eqnarray}
otherwise $B$ would be a decreasing function of $p$, which is
uncorrect. Therefore, this second transition line is limited to the
range $q_2 \ge q_{cr}$. Notice that, as long as $q_1$ is nonzero, the
passage from BP$_2$ to BP$_1$ always involves the crossing of the
boundary between the two percolating phases, i.e., a discontinuous
transition. As we shall see, this is a peculiar feature of the
infinite connectivity limit.

\end{itemize}

\subsection{Large connectivity}

The phase behavior for large $k$ can be obtained by following closely
the steps outlined above for the infinite connectivity case.  We first
approximate
\begin{eqnarray}
 \frac{1}{p} = {\mathcal F}(B) \simeq \frac{1}{ 2 B } \sum_{i=1}^2
 q_i \, \erfc \, Y_{\alpha_i} ,
  \label{F_large_k}
\end{eqnarray}
and setting ${\mathcal F}'(B)=0$ we get:
\begin{eqnarray}
\sum_{i=1}^2 q_i \left[ \erfc \, Y_{\alpha_i} +
 \frac{2 B}{\sqrt{\pi}}  Y'_{\alpha_i} {\rm e}^{ - Y_{\alpha_i}^2 } \right] = 0.
\label{F1_large_k}
\end{eqnarray} 
We then distinguish two regions in which either $B \ge \alpha_1 \ge
\alpha_2$ or $\alpha_1 \ge B \ge \alpha_2$ and observe that when
$\alpha_1$ and $\alpha_2$ are not too close to each other, the
Eq.~(\ref{F1_large_k}) is essentially equivalent to two independent
equations corresponding to the phase boundaries in which either $B
\approx \alpha_1$ or $B \approx \alpha_2$.

In the former case, $B \approx \alpha_1$, we have $\erfc \,
Y_{\alpha_1}(B) \simeq \erfc \, Y_{\alpha_2}(B) \simeq 2$ and
neglecting the term containing ${\rm e}^{ - Y_{\alpha_2}^2} $ in
Eq.(~\ref{F1_large_k}), leads to:
\begin{eqnarray}
  1 & \simeq & q_1 \frac{B}{\sqrt{\pi}} \, Y'_{\alpha_1} {\rm e}^{ -
    Y_{\alpha_1}^2 } .
\end{eqnarray} 
Thus, the first transition line delimiting the BP$_1$ phase is:
\begin{eqnarray}
  p_{\scriptstyle \rm c_1} & = & \frac{2 B_{\scriptstyle \rm c_1} }{
    q_1 \, \erfc \, Y_{\alpha_1} (B_{\scriptstyle \rm c_1}) + q_2 \,
    \erfc \, Y_{\alpha_2} (B_{\scriptstyle \rm c_1}) } ,
\label{eq.pc1}
\end{eqnarray}
with
\begin{eqnarray}
  B_{\scriptstyle \rm c_1} & \simeq & \alpha_1 - \sqrt{
    \frac{\alpha_1(1-\alpha_1) }{k} \ln{ \frac{q_1^2 \, k \,
        \alpha_1}{2 \pi (1-\alpha_1) } } }.
\label{eq.Bc1}
\end{eqnarray}

In the latter case, the phase boundary corresponds to $ B\approx
\alpha_2$ and we can consider $\erfc \, Y_{\alpha_2} \ll \erfc \,
Y_{\alpha_1} \simeq 2$ in Eq.~(\ref{F1_large_k}). Neglecting the term
containing ${\rm e}^{ - Y_{\alpha_1}^2} $ we get
\begin{eqnarray}
  2 & = & \frac{2 B}{\sqrt{\pi}} \, Y'_{\alpha_2} {\rm e}^{ -
    Y_{\alpha_2}^2 } .
\end{eqnarray} 
This leads to the second transition line:
\begin{eqnarray}
  p_{\scriptstyle \rm c_2} & = & \frac{2 B_{\scriptstyle \rm c_2} }{
    q_2 \, \erfc \, Y_{\alpha_2} (B_{\scriptstyle \rm c_2} ) } .
\end{eqnarray}
where $ B_{\scriptstyle \rm c_2} = \Bc(\alpha_2,k)$, with $\Bc$ being
the function defined by Eq.~(\ref{eq.Bc}).

\begin{figure}[htbp]
\begin{center}
\input{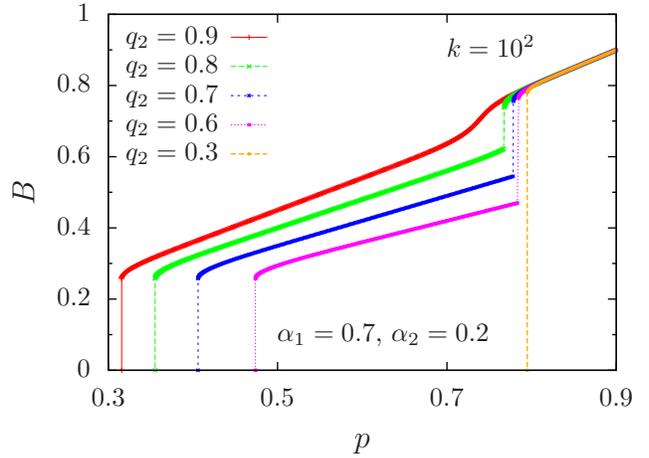}
\caption{Order parameter $B$ vs the occupation probability $p$. For
  $q_2=0.1$ the system passes very near to the ${\mathsf A}_3$
  singularity, while for larger values of $q_2$ it goes through the
  three phases and so the curves show a double jump.}
\label{fig.B_gg_k1e2alpha0308}
\end{center}
\end{figure}

Therefore, also for large connectivity we obtain two transition lines
delimiting two percolating phases, and thus the possibility of
multiple discontinuous phase transitions, see Fig.~\ref{fig.phase_gg}a
for two examples with $k=10^2,\, 10^3$ (for fixed $\alpha_1=0.7$ and
$\alpha_2=0.2$). Note, that also in this two-specie case the relevant
scaling variables on which the two transition lines do depend are $k
\alpha_i/(1-\alpha_i)$ with $i=1,2$.

It is now interesting to observe that, at variance with the $k=\infty$
case, the transition line $ p_{\scriptstyle \rm c_1} $ has a terminal
endpoint corresponding to the vanishing of the logarithm term in
Eq.~(\ref{eq.Bc1}), that is:
\begin{eqnarray}
  q_2^{{\mathsf A}_3} & = & 1 - \sqrt{ \frac{2 \pi (1-\alpha_1)}{k \,
      \alpha_1} }
\label{eq.q_A3}
\end{eqnarray}
This is essentially equivalent, to the leading order in $k$ we
consider, to the point where ${\mathcal F}''=0$, i.e., to an ${\mathsf
  A}_3$ singularity or cusp bifurcation, see Fig.~\ref{fig.phase_gg}a.
The strength of the discontinuous transition between the two
percolating phases weakens as the ${\mathsf A}_3$ critical endpoint is
approached upon increasing $q_2$ until the distinction between the
BP$_1$ and BP$_2$ phases disappears. More generally, the presence of
the critical endpoint implies that there is always a path in the
control variables space which allows a smooth passage between the two
percolating phases, i.e., without observing an extra jump in the order
parameter. In Fig.~\ref{fig.phase_gg}a) we also show the bifurcation
line of ${\mathsf A}_3$ singularities:
\begin{eqnarray}
  p^{{\mathsf A}_3} (q_2) & = & \frac{2 \, \alpha_1 }{ 1-q_2 + q_2 \,
    \erfc \, \frac{ \sqrt{\pi}(\alpha_2 - \alpha_1) }{ (1-q_2) \,
      \alpha_1} }.
\label{eq.p_A3}
\end{eqnarray}
This is obtained by expressing $k$ in terms of $q_2$ through
Eq.~(\ref{eq.q_A3}) and plugging the obtained result in
Eq.~(\ref{eq.pc1}). 

Fig.~\ref{fig.phase_gg}b) instead shows a section of the phase diagram
for several values of $\alpha_1$ at fixed connectivity $k=10^3$ and
$\alpha_2=0.2$.  The topology of the phase diagram remains essentially
unchanged. In particular the ${\mathsf A}_3$ bifurcation line is now
obtained by using $\alpha_1(q_2)$ from Eq.~(\ref{eq.q_A3}) in
Eq.~(\ref{eq.pc1}), and there is no need to report the explicit
expression. Howewer, it is interesting to note that when $\alpha_1$
gets closer to $\alpha_2$ the transition line between the percolating
phases shrinks until a special value of $\alpha_1$ (generally
different from $\alpha_2$) is reached at which only one percolating
phase survives. This occurs when the ${\mathsf A}_3$ critical endpoint
coalesces with the crossover point between the $ p_{\scriptstyle \rm
  c_1}$ and $p_{\scriptstyle \rm c_2}$ transition lines, or,
equivalently, when $p^{{\mathsf A}_3}= p_{\scriptstyle \rm c_2}$. This
corresponds to the formation of an ${\mathsf A}_4$ singularity, also
known as swallowtail bifurcation. In Fig.~\ref{fig.phase_gg}b) the
terminal point of the ${\mathsf A}_3$ cusp line represents the
${\mathsf A}_4$ singularity to the leading order approximation
considered here, which is almost indistinguishable from the exact
result, as we generally find in the large connectivity limit.  An
illustration of the complex phase behavior discussed above is shown in
Fig.~\ref{fig.B_gg_k1e2alpha0308} where we see the typical evolution
of the order parameter near the ${\mathsf A}_3$ singularity and the
double jump it undergoes when the BP$_1$ and BP$_2$ phases are crossed
upon changing $p$.

\section{Conclusions}

To summarize, the large connectivity limit of bootstrap percolation
provides a suitable framework in which one can investigate
analytically the dependence of critical properties on the threshold
parameter and lattice connectivity. We generally found that all the
interesting features of the phase structure found in low connectivity
random graphs, such as hybrid phase behavior, multiple transitions and
higher-order singularities, are robust in this limit, although,
perhaps surprisingly, the infinite connectivity limit turns out to be
singular.

For general $n$-ary mixtures one can reasonably speculate that the
phase structure will entail $n$ percolating phases, provided that the
values of the random variable $\alpha$ (that is, the ratio of
threshold parameter to connectivity) are sparse enough.  As the
initial density of particles $p$ increases, one would then observe, in
a suitable domain of the control variables, a sequence of multiple
transitions starting from a loose percolating phase (the one with the
lowest value of $\alpha$), towards more and more compact percolating
phases (those with higher and higher values of $\alpha$); in
particular, when the lowest value of $\alpha$ is zero the incipient
percolation cluster that first appears has a fractal structure and the
associated phase transition is an ordinary random percolation
transition.

We expect that the present framework is especially relevant for the
study of massively connected networks and that it can be rather
effective in the study of threshold dynamics closely related to
BP~\cite{Altarelli}.  We also remark that the large
connectivity limit could be a valuable tool for addressing the
dynamics of facilitated spin models of glasses which, in spite of
their apparent simplicity, still lacks an analytic solution. Work in
this direction is in progress.


\begin{thebibliography}{50}
\bibitem{NB} For simplicity, we depart here from the standard
  definition of the order parameter as the probability $P$ that a site
  belongs to the infinite cluster.  In fact, since $P$ is a polinomial
  function of $B$, $$ P = p \sum_{i=0}^{z-m} {z \choose i} B^{z-i}
  (1-B)^i, $$ its critical properties near the transition are the same
  as those of $B$.
\end{thebibliography}
\end{document}